\documentclass[10]{article}
\pdfoutput=1
\usepackage[top=1in,bottom=1in,left=1in,right=1in]{geometry}
\usepackage{graphicx}
\usepackage{biblatex}
\usepackage{amssymb}
\usepackage{amsmath}
\usepackage{mhchem}
\usepackage{gensymb}
\usepackage{multicol}
\usepackage{caption}
\usepackage{indentfirst}
\usepackage{hyperref}
\usepackage{lscape}
\usepackage{tabularx}
\usepackage{mathtools}
\usepackage[singlelinecheck=false]{caption}
\usepackage[bottom]{footmisc}

\usepackage{placeins}
\usepackage{caption}

\usepackage{lineno}

\usepackage{wrapfig}
\usepackage{lipsum}

\usepackage{textcomp}
\usepackage[hyphenbreaks]{breakurl}

\usepackage{ragged2e}

\begin{document}
\title{\textbf{Thermal conductivity measurements of PTFE and Al$_{2}$O$_{3}$ ceramic  at sub-Kelvin temperatures}}
\author{Alexey Drobizhev$^{\mathrm{a},\mathrm{b},*}$, Jared Reiten$^{\mathrm{a}}$, Vivek Singh$^{\mathrm{a},\mathrm{b}}$, Yury G. Kolomensky$^{\mathrm{a},\mathrm{b}}$ \\
\footnotesize a -- \textit{Department of Physics, University of California, Berkeley, 366 LeConte Hall, Berkeley, CA 94720, USA} \\
\footnotesize b -- \textit{Nuclear Science Division, Lawrence Berkeley National Laboratory, 1 Cyclotron Rd, Berkeley, CA 94720, USA }
\\ 
\vspace{1mm}
} 
\date{7 December 2016}
\maketitle

{\footnotesize  The design of low temperature bolometric detectors for rare event searches necessitates careful selection and characterization of structural materials based on their thermal properties. We measure the thermal conductivities of polytetrafluoroethylene (PTFE) and Al$_2$O$_3$ ceramic (alumina) in the temperature ranges of 0.17--0.43 K and 0.1--1.3 K, respectively. For the former, we observe a quadratic temperature dependence across the entire measured range. For the latter, we see a cubic dependence on temperature above ~0.3 K, with a linear contribution below that temperature. This paper presents our measurement techniques, results, and theoretical discussions.

}

\begin{multicols}{2}

\section{Introduction}

Cryogenic bolometers used as particle detectors read out thermal signatures of events in a cold absorber crystal, and offer several advantages for rare event physics searches. The most notable of these is the excellent energy resolution, which is very useful for isolating and identifying phenomena like neutrinoless double beta decay and dark matter interactions [1]. Furthermore, a bolometer's absorber can be fabricated from a wide range of materials, including those that incorporate radiation sources. This allows for good scalability, exemplified by the upcoming CUORE (Cryogenic Underground Observatory for Rare Events) ton-scale experiment, which features 988 crystals/channels of TeO$_2$, containing the neutrinoless double beta decay candidate isotope  $^{130}$Te [2]. 

Due to the very low operating temperatures of these devices (10--100 mK for state of the art examples), their design and operation also present unique technical challenges. Among these are the characterization and selection of appropriate structural materials. 

The bolometer's absorber is coupled by a weak link to a heat sink. The weak link's thermal conductivity $K(T)$, needs to be low, so that event heat energy is not immediately dissipated to the bath and to cause a temperature rise in the absorber that is high and long enough to make a good measurement. These considerations, combined with wide availability, easy machining, and good radiopurity motivates the widespread use of PTFE---(C$_2$F$_4$)$_{\mathrm{n}}$, a soft plastic commonly known by the trademark name Teflon---in many instruments, including CUORE and Cuoricino [2][3]. Given its popularity in cryogenics, low temperature thermal properties of PTFE have been studied extensively. The thermal conductivity has been measured down to 170 mK, revealing an approximately quadratic temperature dependence that is consistent with theoretical models of amorphous solids [4]. However, the parameter that truly optimizes a weak link is thermal diffusivity,
\begin{equation}
\alpha(T) = K(T)/C(T),
\end{equation}
through which heat capacity $C(T)$ also comes into play. While we want the weak link to have high impedance to heat dissipation from the absorber, it should not store heat itself. Such storage causes the undesirable effects of lengthening the bolometer pulses when the stored heat diffuses back into the absorber and sensor, and decreases their amplitude when a large portion of the heat dissipates directly to the bath. PTFE's high heat capacity compared to typical bolometric absorber materials [5] makes it less than optimal in this role, especially since its low structural strength necessitates relatively large parts. Efforts at constructing a thermal model of the CUORE bolometers in the form of a numerically evaluated system of classical differential equations reveal the strong effect PTFE's low diffusivity has on pulse shape [6].

Alumina interests us, because it has low heat capacity and strong mechanical rigidity. It is prepared by sintering (compacting under high pressure and temperature) corundum/sapphire crystallites. Use of alumina in cryogenic experiments is currently less extensive than PTFE, as is the characterization of its thermal properties at ultra-low temperatures. Measurements of its thermal conductivity exist down to $\sim$ 1 K and higher. These exhibit approximately quadratic or cubic temperature dependencies, as well as wide sample-to-sample variation in absolute value due to manufacturing differences [7][8].  We measure the thermal conductivity of alumina in the $\sim$ 0.1--1.3 K range, and PTFE in the 0.17--0.43 K range. Our PTFE  measurements are carried out at significantly finer temperature increments than currently available results, useful for our efforts at bolometer thermal modeling, as well as validating our experimental technique for the more poorly understood alumina.

\section{Experimental setup}

\subsection{Technique}

We perform our measurements using the ``integrated thermal conductivity method'' [9]. This involves applying known powers $P_{\mathrm{applied}}$ on one end of the sample and measuring that end's temperature $T$, while maintaining the other end at a known constant base temperature $T_0$.

The total thermal power $P(T)$ applied to the sample is equal to the integral of the thermal conductivity of the sample over the measured temperature gradient multiplied by a geometric factor
\begin{equation}
P(T) = A/l\times\int^T_{T_0}K(T')dT',
\end{equation}
where $A$ is the cross-sectional area of the sample, $l$ is its length, and $K(T)$ is thermal conductivity. Notice that maintaining $T_0$ constant makes the thermal power a function of the upper limit of the integral, simplifying analysis. 

Ideally, there would be no gradient at all across the sample while the heater is switched off. However, in practice we often observe a parasitic heat load. Sources can include leakage current from the heater power supply, thermal radiation from the sample's surroundings, vibrational noise, and RF noise/pickup. Making the assumption that this unwanted contribution is constant, we include it as an additional parasitic power, $P_{\mathrm{parasitic}}$:
\begin{equation}
P(T) = P_{\mathrm{applied}}(T) + P_{\mathrm{parasitic}}.
\end{equation}
The presence of a constant offset does not affect our analysis, since we extract thermal conductivity by differentiation, which negates the effect of a constant parasitic heat load:
\begin{equation}
K(T) = l/A\times\frac{dP(T)}{dT} = l/A\times\frac{dP_{\mathrm{applied}}(T)}{dT}.
\end{equation}

We perform the above calculation by fitting our $P_{\mathrm{applied}}(T)$ data with a power law plus a negative constant offset for $P_{\mathrm{parasitic}}$ (separately for each sample and dataset), and differentiating the result analytically. Uncertainties are propagated through.

\subsection{Hardware}

The low temperature conditions are achieved using an Oxford Instruments Triton 400 dilution refrigerator. This cryostat combines a standard $^3$He-$^4$He dilution unit with a ``cryogen-free'' pulse tube cooler setup to reach temperatures as low as 6 mK. Nominal cooling power at 100 mK is 400 $\mu$W. In practice, for these measurements temperatures are in the hundreds of millikelvins due to the heat loads involved.

To create the temperature gradients needed to measure thermal conductivity, one end of an elongated sample is thermalized to the mixing chamber plate (coldest stage) of the cryostat, while the other end hangs freely and is equipped with a heater. A pair of thermometers measure the temperatures of both ends.

Four 3'' (7.6 cm) long alumina samples are measured: a pair of cylindrical rods 0.25" (6.35 mm) in diameter (henceforth referred to as samples A and B), a prism with a 0.25"$\times$0.25" (6.35$\times$6.35 mm$^2$) square cross section (sample C), and a thicker cylinder 0.75'' (1.9 cm) in diameter (sample D). All of these samples are ``Nonporous Alumina Ceramic Rods'' acquired from McMaster Carr, with a quoted porosity of 0\% and purity of 96--99.8\% Al$_2$O$_3$ [11]. The first PTFE sample (sample 1) used is a 2.75'' (7.0 cm) long rectangular prism with a 0.25"$\times$0.25" (6.35$\times$6.35 mm$^2$) square cross section. The second (sample 2) was cut from the same brick of PTFE as the first, with the same cross-sectional area and about half the length. The PTFE was also acquired from McMaster Carr, under the name ``Ultra-Clean PTFE'' [11].

The thermometers used for all of these measurements are Lakeshore Cryotronics RX-102B-CB RuO$_2$ devices (numbers 2 and 9 in Figures 1--3). These temperature sensors are sensitive within our entire measurement range. We calibrate them in separate cooldowns, in reference to a factory-calibrated RuO$_2$ thermometer and a $^{60}$Co nuclear orientation thermometer. The factory thermometer mounts are made from copper, and are bolted to surfaces. Readout of the RuO$_2$ thermometers is performed with a Lakeshore AC370 resistance bridge. This instrument also includes a PID (Proportional, Integral, Differential) controller, which is used with the factory mixing chamber plate heater and thermometer, also a RuO$_2$ (not seen in figures), to stabilize the temperature of the cold bath. 

We use two heater types. In our first measurements, with alumina samples A-C, these are Lakeshore Cryotronics 50 $\Omega$ cartridge heaters (wound NiCr wire, no. 10 in Figure 1). We drive these with a Sorensen LS 18-5 DC voltage source. Given the low resistance of the cartridge heaters, a room temperature load resistor is added in series. The resulting setup suffers from significant unwanted heating from leakage current, as can be seen in the large parasitic offset of those three datasets in Figure 8. For the thicker alumina sample D and PTFE, we upgrade to 50 k$\Omega$ (alumina) and 1 k$\Omega$ (PTFE) NiCr thin film heaters (no. 10 in Figures 2 and 3). These are in the form of chips varnished to a small Kapton flexible PCB, with gold wire bonds forming electrical connections to a pair of solder pads. For biasing, we switch to the built-in DC voltage source of a Keithley 6517B  electrometer for alumina, and a Keithley 6220 precision DC current source for PTFE. Given the higher resistance of the thin film heaters, no additional load is used. This arrangement provides significantly lower leakage and parasitic heating, as manifested by the smaller offsets in Figures 4 and 8.

Three different variations of sample holder design are used for alumina samples A-C, alumina sample D, and the PTFE (Figures 1--3). In the case of alumina, the ends of the sample are firmly held in vise-like copper clamps with cutouts in the jaws in the shape and size of the sample cross section. The top clamp, for the cold end, is flat with a large surface area for secure mounting and thermalization in the cryostat (no. 4 in Figures 1 and 2). For samples A-C, the bottom clamp is brick-like, covering the sample from three sides. Below the sample is a 0.25" (6.35 mm) diameter channel perpendicular to the sample, exactly matching the cartridge heater's dimensions, in which it is embedded with Dow Corning high vacuum grease impregnated with copper powder for improved thermal coupling (Figure 1, nos. 8 and 10). The RuO$_2$ thermometer is attached to the clamp (Figure 1, no. 9). For sample D, the bottom clamp is also a flat design more similar to the top holder (Figure 2, no. 8). The thermometer (no. 9) is bolted to its top surface, and the thin film heater (no. 10) is adhered with GE varnish. Since the heater now directly contacts only one of two jaws, a flexible sheet of copper (no. 11) 

\columnbreak

{\centering
\includegraphics[width=0.75\columnwidth]{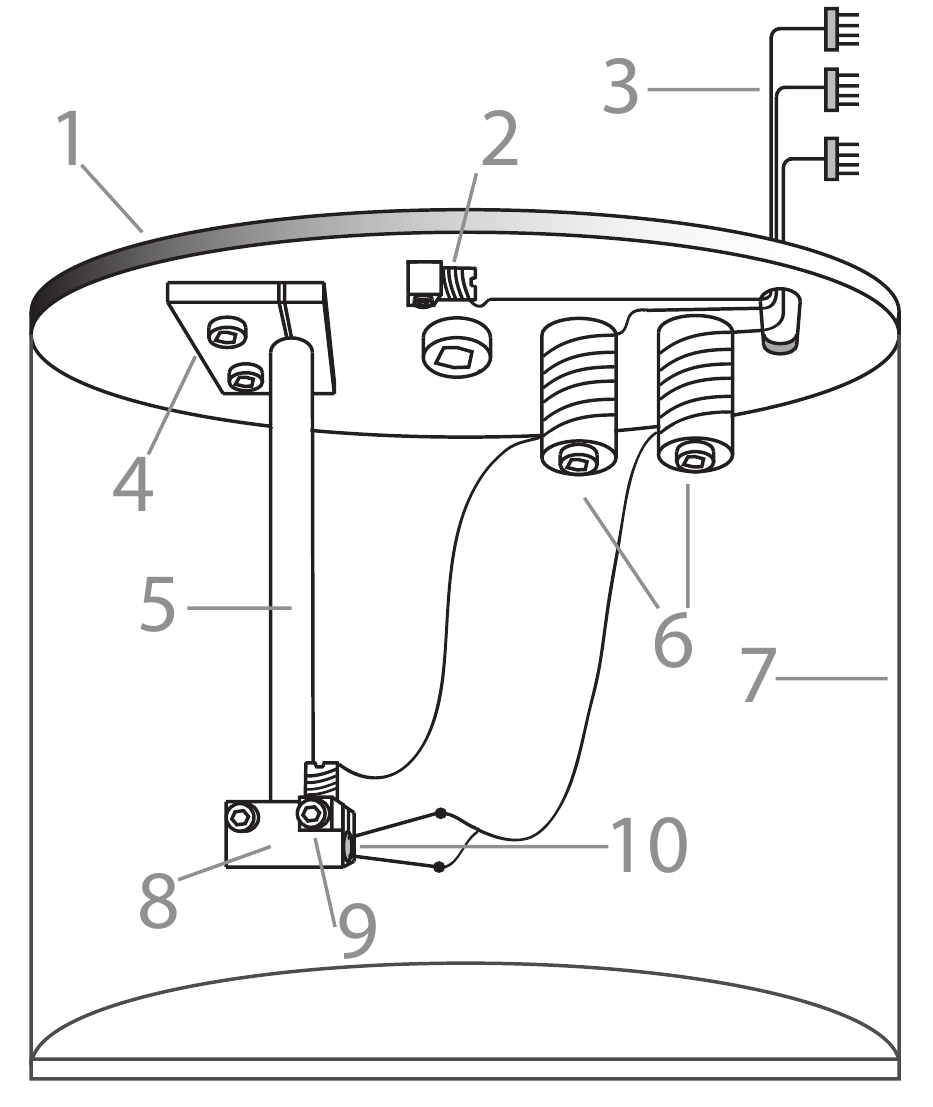}\\
\captionof{figure}{\footnotesize{Experimental setup for the 0.25" diameter cylindrical and 0.25"$\times$0.25" square alumina samples. 1--Cu plate, thermalized to mixing chamber; 2--cold end sensor; 3--electrical connections; 4--cold end Cu clamp; 5--sample; 6--thermalization bobbins; 7--Cu shield; 8--heated end Cu clamp; 9--heated end sensor; 10--cartridge heater (embedded).}}\label{pinki}
}

\vspace{2mm}

{\centering
\includegraphics[width=0.75\columnwidth]{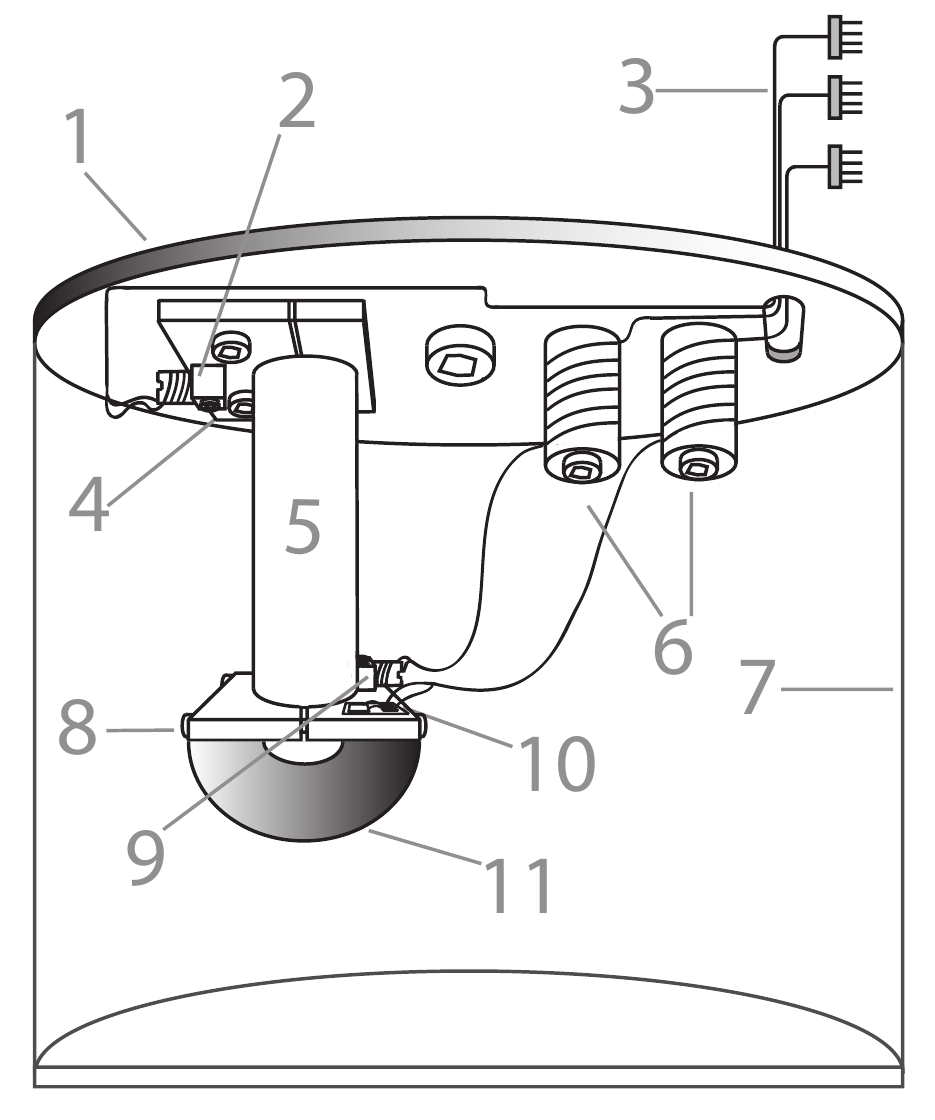}\\
\captionof{figure}{\footnotesize{Experimental setup for the 0.75" diameter cylindrical alumina sample. 1--Cu plate, thermalized to mixing chamber; 2--cold end sensor; 3--electrical connections; 4--cold end Cu clamp; 5--sample; 6--thermalization bobbins; 7--Cu shield; 8--heated end Cu clamp; 9--heated end sensor; 10--thin film heater; 11--Cu sheet thermalizing clamp jaws.}}\label{pinki}
} 

\vspace{2mm}

\noindent joins the two jaws of the clamp to ensure even thermalization. The top clamps are generally similar, though for the thick sample the cold end RuO$_2$ sensor (no. 2 in Figures 1 and 2) is moved directly to the clamp from its original position nearby. In all cases, samples penetrate the clamps to a depth of 0.25'' (6.35 mm). The length of the sample is defined as the distance between the clamp edges, and does not include the embedded portions of the rod. \\
\indent We find the clamp design to be inadequate for PTFE---differential thermal contraction, combined with the material's soft and slippery texture, causes the holders to separate from the sample during cooling, even when clamps are upgraded with flexible leaf springs. Instead, we drill holes in the samples and use brass threaded rods to make attachments, without thermal grease (Figure 3). On the warm end, besides the RuO$_2$ and the thin film heater on a small copper rectangle on opposite sides of the sample, we attach two copper bricks on the remaining two sides. These are both connected to the heater with copper wire for more even heat distribution. The cold end setup is analogous, except the heater is replaced by an attachment to a T-shaped thermalized base. Of all the elements, the two thermometers are mounted closest to the center of the sample. The length of the PTFE sample is defined by the distance between these. Mounting the thermometer separately from the heater has an additional advantage in reducing difficult-to-characterize parasitic heat load and holder-on-sample thermal boundary resistance [10].

\vspace{5mm}
{\centering
\includegraphics[width=0.75\columnwidth]{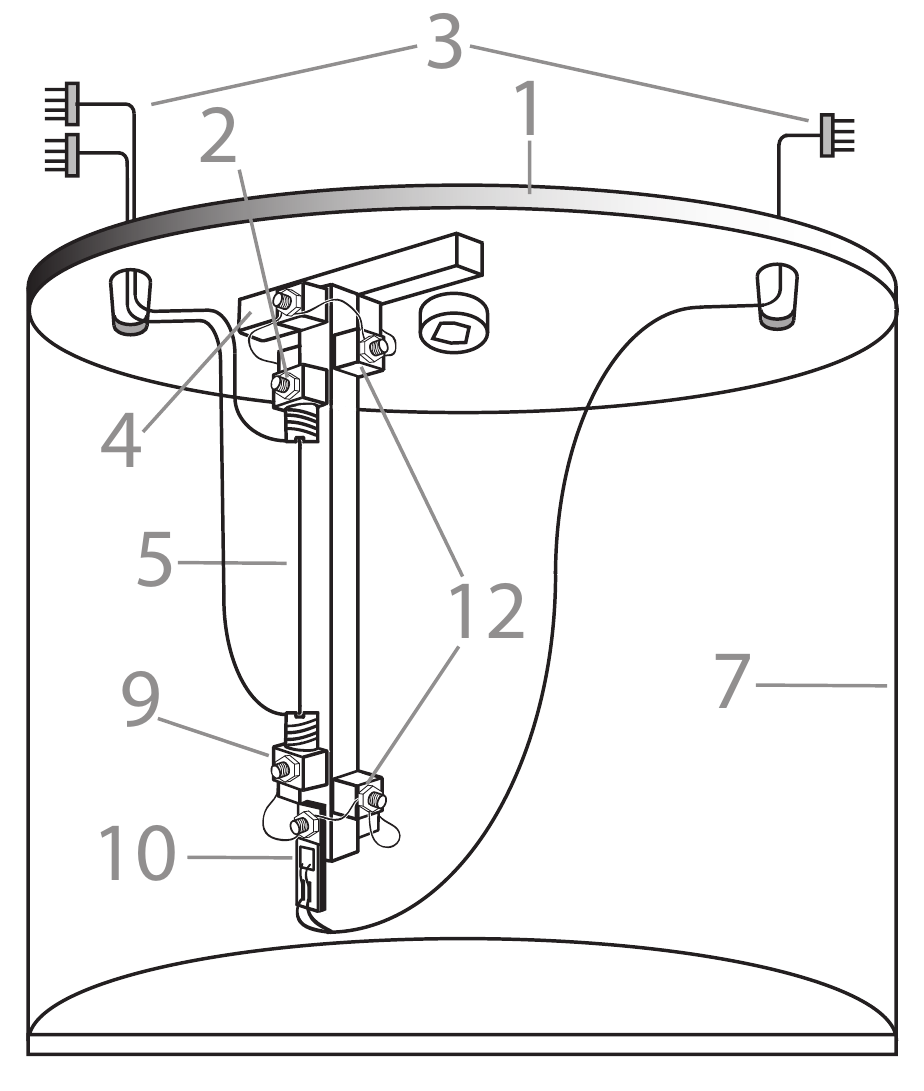}\\ 
\captionof{figure}{\footnotesize{Experimental setup for the PTFE samples, featuring threaded rod attachment of sensors and heater. 1--Cu plate, thermalized to mixing chamber; 2--cold end sensor; 3--electrical connections; 4--Cu cold end base; 5--sample; 7--Cu shield; 9--heated end sensor; 10--thin film heater; 12--thermalization bricks (3 on each end, connected to base/heater by Cu wire). Wire thermalization bobbins were omitted, as they were found to be unnecessary.}}\label{pinki}
}

\vfill
\columnbreak

\subsection{Shunt conductance, thermal boundary resistance, and parasitic heat load}

Technically, we are not measuring the thermal conductance of the sample on its own. Instead, we measure the effective conductance of a system consisting of the sample connected in series with the copper-on-sample thermal boundary resistance, and in parallel with a ``shunt'' in the form of the thermometer and heater wiring. The thermometer connections are Manganin alloy, which has electrical resistance nearly independent of temperature. Its thermal conductivity is measured to be about $5\times10^{-3}$ -- $8\times10^{-2}$ W/m/K in the 0.1--1.0 K range [12]. Heaters are connected with superconducting NbTi wiring, which conducts heat less---about $1\times10^{-4}$ -- $2\times10^{-2}$ W/m/K in the same range [13]. All wires are 15 -- 20 cm long and 36 AWG. Looking at these properties and at our final results (Figures 5 and 9), we see that the shunt conductance can be neglected.

Determining thermal boundary resistance between the holder and sample is more complicated---it depends on temperature, power applied across the boundary, and material-specific properties [10]. Lack of knowledge of these properties prevents us from doing a precise theoretical calculation, whereas a fully experimental characterization is out of the reach of this study. We use measured and calculated values for similar materials taken from [14] and [15] to make conservative estimates of this contribution. For alumina, we expect a maximum possible boundary temperature gradient of $\sim$ 0.05 mK at 100 mK, where the total gradient is more than $\sim$ 12 mK. At 1 K, the boundary contribution is no greater than $\sim$ 0.5 mK, while total gradient is in the hundreds of millikelvins.  With PTFE, we compute a boundary effect resulting in a maximum gradient of $\sim$ 0.6 mK at 170 mK, where the total gradient is $\sim$ 35 mK, and a boundary contribution of $\sim$ 0.8 mK at 430 mK, where the total gradient is $\sim$ 80 mK. This suggests that boundary effects are not significant, and we include their contributions in our systematics.

Besides intentional heating, we experience a parasitic heat load on our samples, the primary sources of which we believe to be leakage current in the sample heater, and a combination of thermal radiation and mechanical vibration. The former is particularly prominent in the datasets of samples A, B, and C, which use the less optimal Sorensen power supply and low resistance cartridge heaters. We confirm its influence by fully unplugging that connection, which eliminates most of the offset. The switch to high resistance thin film heaters driven by Keithleys for the later measurements significantly reduces the effect. Radiative heating comes from black body radiation from the sample's surroundings. Our cryostat's innermost thermal shield is at the 600--700 mK level. To decrease the radiation heat load on the experimental setup, the hardware is mounted inside an additional copper shield thermalized to the cryostat's mixing chamber plate (Figures 1--3). That said, the shield is fabricated from OFHC (oxygen free, high conductivity) copper, which is less than ideal for cryogenic use due to high hydrogen content, and is of large mass. Furthermore, it is exposed directly to the still vessel's radiation---our cryostat is unable to host a shield on the ``100 mK'' stage. Thus, equilibrium base (mixing chamber) temperature is on the order of 100 mK, and the resultant heat load dominates the parasitic power we observe in the runs instrumented with thin film heaters and Keithleys. A simple calculation with the Stefan-Boltzmann law is consistent with this. The reason the radiation results in a gradient on the sample is that, while impingement on all surfaces is more or less even, the low thermal conductivity of the sample means that any heat absorbed by it and its hot-end hardware is dissipated much more slowly than that absorbed by the bath. Vibrational heating contributes an unknown amount of heat that is ``mixed in'' with the radiative component. Given the consistency of our observed conditions with black body emission and absorption, such a mechanical contribution would be of comparable magnitude or less than radiation. Though we cannot decouple the effect of vibrational heating, it is not necessary to extract thermal conductivity, since the system experiences no time-varying impacts or stresses and it would likewise be approximately constant.

The leakage current from the supply is constant while it is on and connected, and the heaters' resistances are found to be temperature-independent and stable within 2\% or less. The temperature of the cold bath and thermal shield is kept stable by PID control, ensuring the radiative load is also constant. Thus, we expect the parasitic heat load to be constant in the temperature range of interest, which is confirmed experimentally. 

Bath base temperatures as low as $\sim$ 70 mK are achieved, but are not possible to maintain for higher heater powers, creating the main constraint for the lower end of our measurement range.

\subsection{Experimental uncertainties}

In our power versus temperature data, we handle statistical and systematic uncertainties separately. For the former, we take the spread of our temperature readings at every nominal power setting. Each spread is actually entangled with correlated errors stemming from thermometer calibration and PID temperature control. Thus, we adjust them with a common scaling factor to get $\chi^2 /\mathrm{df} = 1$ during fitting and take covariances into account in propagation. The systematic uncertainty on applied power derives primarily from heater
\footnote{Heaters are wired in a four-wire configuration and characterized in separate cooldowns.}
and load resistances, and voltage/current uncertainties from the power supply. Systematic temperature errors come from the thermometer calibrations, as well as a contribution from thermal boundary resistance estimates. In Figures 4 and 8, the statistical uncertainties are indicated by horizontal error bars, while the shaded bands represent systematics. The bands are defined by combining temperature and power uncertainties (conservatively assuming 100\% correlation), shifting the data points by the appropriate amount, and repeating the fit. As with statistics, covariances are included in propagation calculations. Error values quoted in the text and tables, as well as the shaded bands in Figures 5 and 9, represent the combination of statistical and systematic components.

\section{Results and discussion}

\subsection{PTFE}

We measure the thermal conductivity of PTFE over four temperature ranges. The first sample is measured over three ranges within the interval of 0.18--0.43 K. The second sample is measured over a narrow temperature range of about 0.17--0.21 K. The lower bounds are determined by the fact that, at temperatures below these, the thermal gradients across the samples are large and unstable in time. We prioritize making high resolution measurement with small temperature increments over covering a wide temperature range, which has already been done in an overlapping domain [4][16]. 

For each data set we perform a shifted power law fit of the form
\begin{equation}
P_{\mathrm{applied}}(T)\times l/A = a\biggl(\frac{T}{1 \; \mathrm{K}}\biggr)^b + c.
\end{equation}
The data is plotted in Figure 4 with the fit results, uncertainties, and correlations given in Table A.1. The thermal conductivity obtained by differentiation is given in Figure 5 and Table 1, with equations of the form 
\begin{equation}
K(T) = \mu \biggl(\frac{T}{1 \; \mathrm{K}}\biggr)^{\nu},
\end{equation}
where $\mu = ab$ and $\nu = b-1$. We observe an approximately quadratic temperature dependence of $K(T)$, and find $\mu$ and $\nu$ to be strongly correlated. We attribute the $\sim 1\sigma$ discontinuity at 0.37 K in Figure 5 to the relatively narrow temperature interval over which we measure.

Since samples 1 and 2 are cut from the same brick of PTFE, they are considered equivalent. We may thus take a weighted average of the data presented in Table 1, with each $\mu$ and $\nu$ weighted by the reciprocal square of their respective error. We yield the result:

\begin{equation}
K_{\mathrm{w}}(T) = (5.0 \pm 0.8)\times 10^{-3} \biggl(\frac{T}{1 \; \mathrm{K}}\biggr)^{1.83\pm 0.11} \; \mathrm{W/m/K}.
\end{equation}

Our findings are consistent with the tunneling model. This theory describes amorphous solid structure by making the assumption that each atom has access to two neighboring equilibrium sites, modeled as a double well potential, within which tunneling occurs [17][18]. It is applied to broadly defined amorphous solids, including plastics, by Phillips [18]. His treatment focuses on the tunneling states' frequency-dependent relaxation time, $\Gamma (\omega)$, to recover the thermal conductivity temperature dependence: 
\begin{equation}
K(T) \propto T^2.
\end{equation}

$\Gamma (\omega)$ is experimentally shown to account for tunneling states in PE (polyethylene, (C$_2$H$_4$)$_{\mathrm{n}}$), which is similar to PTFE ((C$_2$F$_4$)$_{\mathrm{n}}$) in molecular and physical structure [18]. Thus, we would expect analogous tunneling states---the structural units providing two equilibrium positions and the differences between hydrogen and fluorine determining the height of the potential barrier separating them. This makes the tunneling model a reasonable theory to account for the thermal conductivity of PTFE at cryogenic temperatures.

T. Scott and M. Giles measure the thermal conductivity of PTFE from 0.17--4 K and observe a strong $T^2$ dependence below 1.2 K [4]. Anderson et al. report a $T^{2.4}$ dependence in the range of 0.3--0.7 K [16]. We observe a temperature dependence, as well

\columnbreak

{\centering
\includegraphics[width=\columnwidth]{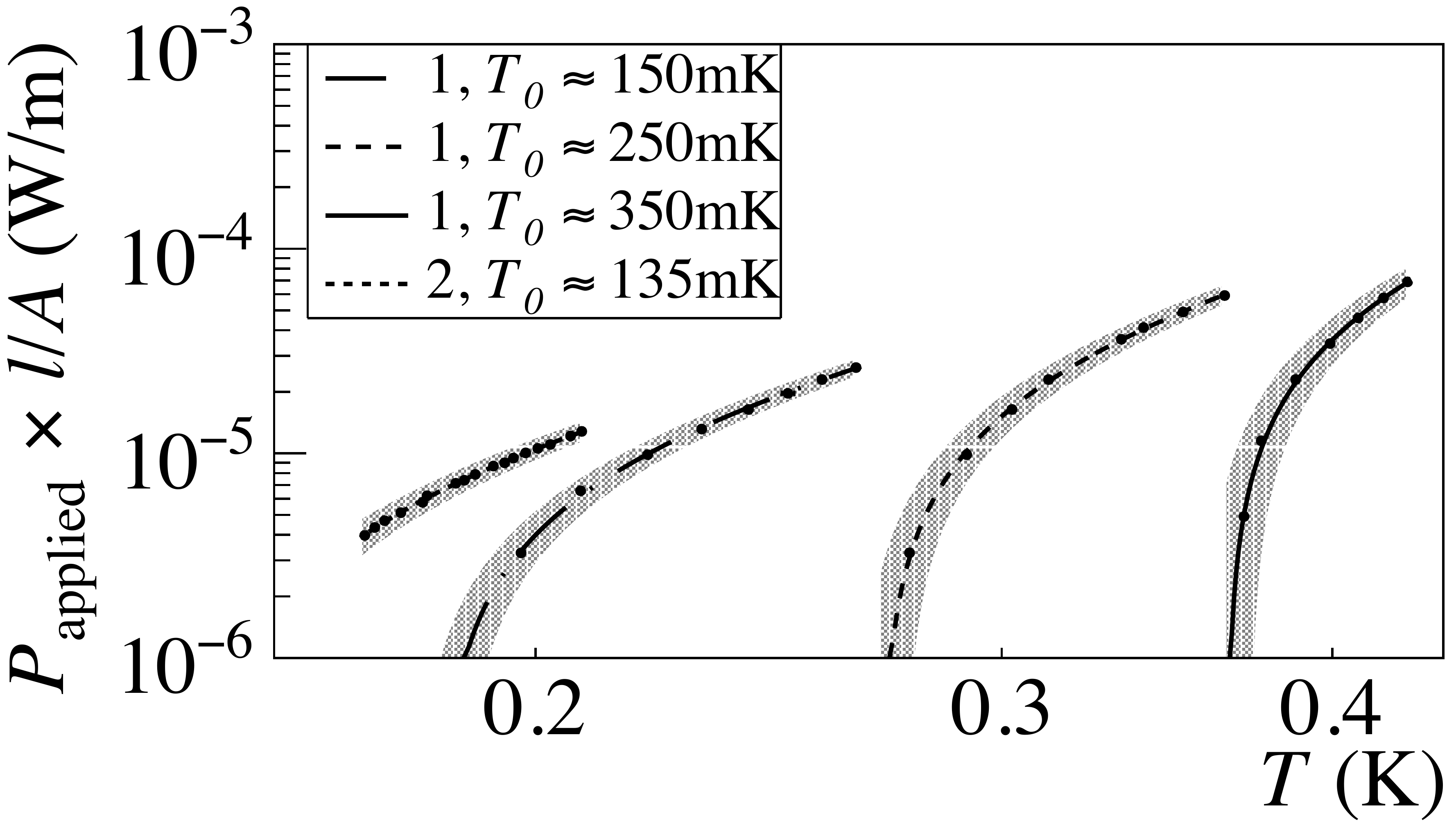} \\
\captionof{figure}{\footnotesize{PTFE measurements: applied heater power scaled by sample length $l$ and cross-sectional area $A$ as a function of heated end temperature $T$. Error bars indicate statistical uncertainty. Bands represent systematics. Numbers in legend refer to samples.}}\label{pinki}
}
\vspace{2mm}

{\centering
\includegraphics[width=\columnwidth]{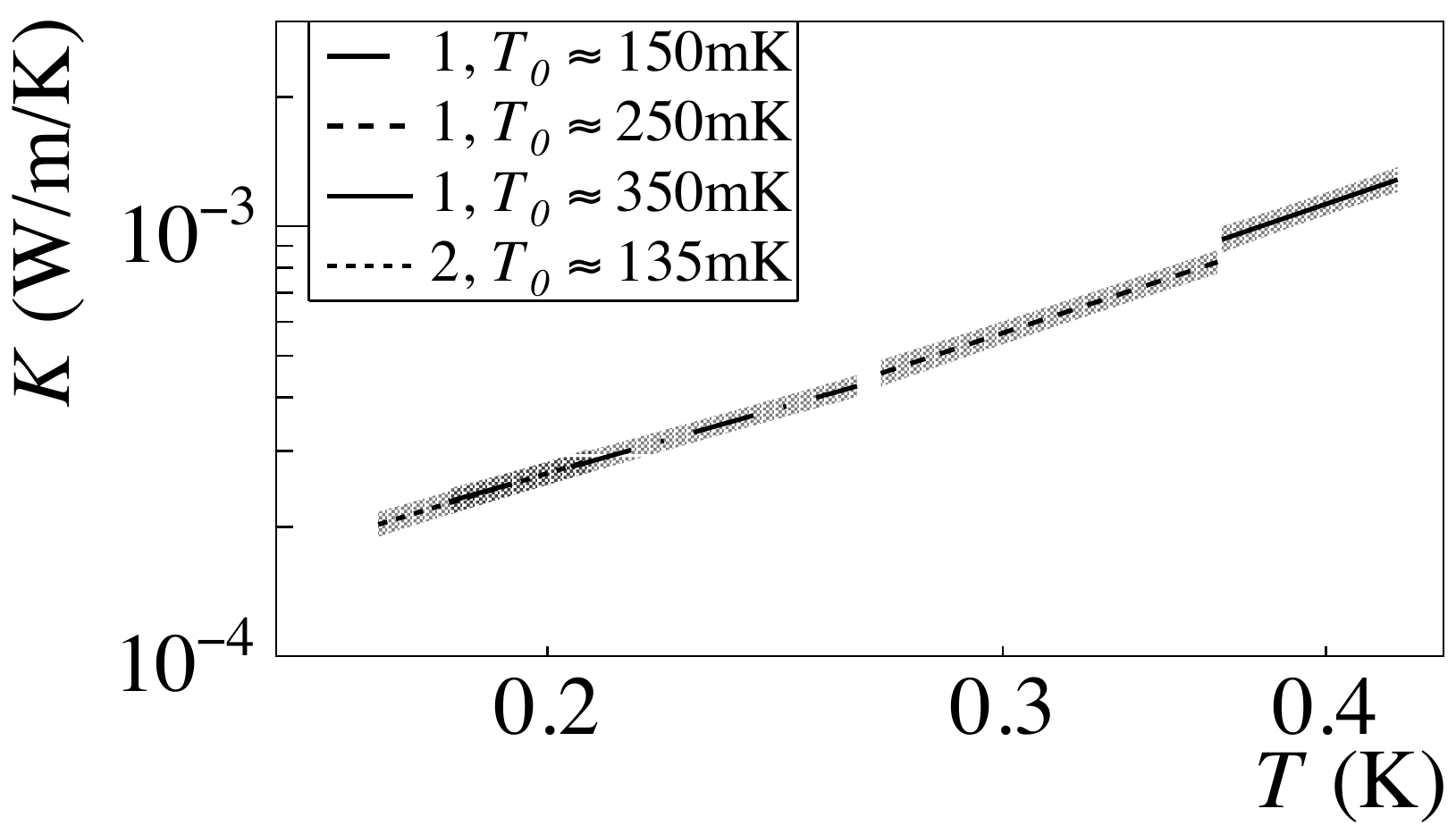} \\
\captionof{figure}{\footnotesize{PTFE thermal conductivity determined by differentiation of scaled $P_{\mathrm{applied}}$ fit in Figure 4. Shaded bands represent combined propagated statistical and systematic uncertainty. Numbers in legend refer to samples.}}\label{pinki}
}
\vspace{2mm}

\noindent as absolute value, of thermal conductivity that agree with those reported by T. Scott and M. Giles within the entire measurement range of 0.17--0.43 K. In Equation 7, the scaling factor is about a 30\% larger than that observed by A.C. Anderson et al. with the temperature power being about 25\% smaller. These  differences result in an absolute value of about a factor of 2.3 times larger than that found by these authors in the overlapping regime of 0.30--0.43 K. We may account for the disagreement through the variability in samples that occurs in the industrial

\end{multicols}
\begin{table}[h]
\centering
{\def\arraystretch{1.5}\tabcolsep=10pt
\scalebox{0.87}
{
\begin{tabular}{c c c c c}
\hline
Sample & Temperature Range (K) & $\mu$ (W/m/K) & $\nu$ & corr[$\mu$,$\nu$] \\
\hline
1 & 0.18--0.26 & (4.1 $\pm$ 1.1)$\times 10^{-3}$ & 1.70 $\pm$ 0.17 & 0.991 \\
1 & 0.27--0.36 & (6.2 $\pm$ 1.5)$\times 10^{-3}$ & 1.99 $\pm$ 0.20 & 0.988 \\
1 & 0.37--0.43 & (7.4 $\pm$ 2.8)$\times 10^{-3}$ & 2.05 $\pm$ 0.41 & 0.995 \\
2 & 0.17--0.21 & (5.0 $\pm$ 1.8)$\times 10^{-3}$ & 1.82 $\pm$ 0.22 & 0.994 \\ \hline

\end{tabular}
}
}
\captionsetup{justification=centering}
\caption{\footnotesize{Results for PTFE thermal conductivity: $K(T)= \mu (T/1 \; \mathrm{K})^{\nu}$.}}
\end{table}

\begin{multicols}{2}

\noindent manufacturing process of PTFE.

\subsection{Alumina}

We acquire four alumina datasets, corresponding to the four samples. For samples A-C, all of which are measured above 0.35 K, we perform a single fit using Equation 5 over the entire data range. The constant $c$ accounts for the parasitic heat load. Equation 5 agrees well with the data. The fits can be seen in Figure 8 and are given in Table A.2, together with parameter uncertainties, correlations, and fit domains. The relatively large negative constant offset, seen prominently in Figure 8, stems mainly from leakage current from the heater power supply used for these measurements.

We measure alumina sample D over a temperature range that is both wider and colder (down to ~0.1 K). A fit of the form of Equation 5 is a visibly poor representation of the data, as can be seen in Figure 6. Examining the graph, there appears to be a kink or transition around 0.3 K, below which the slope is more shallow. Thus, we fix $b$ in Equation 5 to the ``crystal'' value of 4 and add second power law term: 
\begin{equation}
P_{\mathrm{applied}}(T)\times l/A = a\biggl(\frac{T}{1 \; \mathrm{K}}\biggr)^4 + r\biggl(\frac{T}{1 \; \mathrm{K}}\biggr)^s + c.
\end{equation}

Figure 7 clearly demonstrates that Equation 9 agrees well with the data, giving a quadratic dependence in the second power law term. 

This fit is given in Figure 8 and Table A.2 together with the results for other samples, with parameter correlations in Matrix A.4. Due to the much cleaner heater power supply, and consequently the lower parasitic heat load, the constant offset is much smaller in the sample D data than for the others. Differentiating all four fits gives alumina thermal conductivity results, plotted in Figure 9 and given in Table 2 with parameter correlations in Matrix A.3. Equation 6 describes samples A-C, while sample D follows an equation of the form
\begin{equation}
K(T) = \mu \biggl(\frac{T}{1 \; \mathrm{K}}\biggr)^3 + \lambda \biggl(\frac{T}{1 \; \mathrm{K}}\biggr)^{\gamma},
\end{equation}
where $\lambda = rs$ and $\gamma = s-1$. 

Above 0.3 K, in all our data sets, we recover a close to cubic dependence of thermal conductivity on temperature, which is consistent with crystalline/crystallite structure. The thermal conductivity of crystalline insulators at temperatures below a few Kelvin is dominated by lattice vibrations, with phonons being the carriers of heat. In this regime, the phonon mean free path, $l$, is dependent on the

\vfill
\columnbreak
{\centering
\includegraphics[width=\columnwidth]{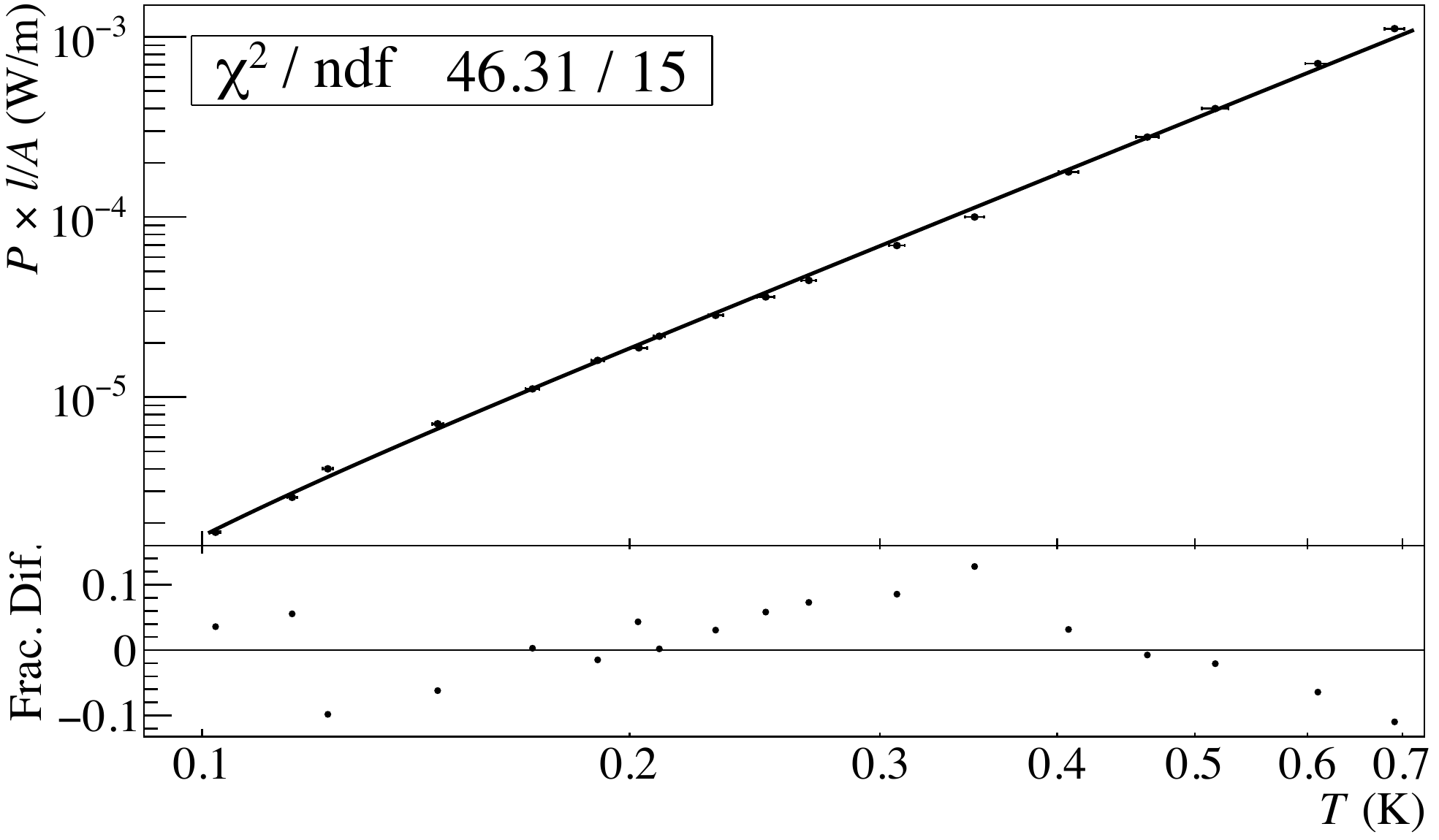}\\
\captionof{figure}{\footnotesize{Alumina sample D data with single power law fit (equation 5). Fractional differences between fit and measured values are plotted below the main graph.}}\label{pinki}
}
\vspace{2mm}

{\centering
\includegraphics[width=\columnwidth]{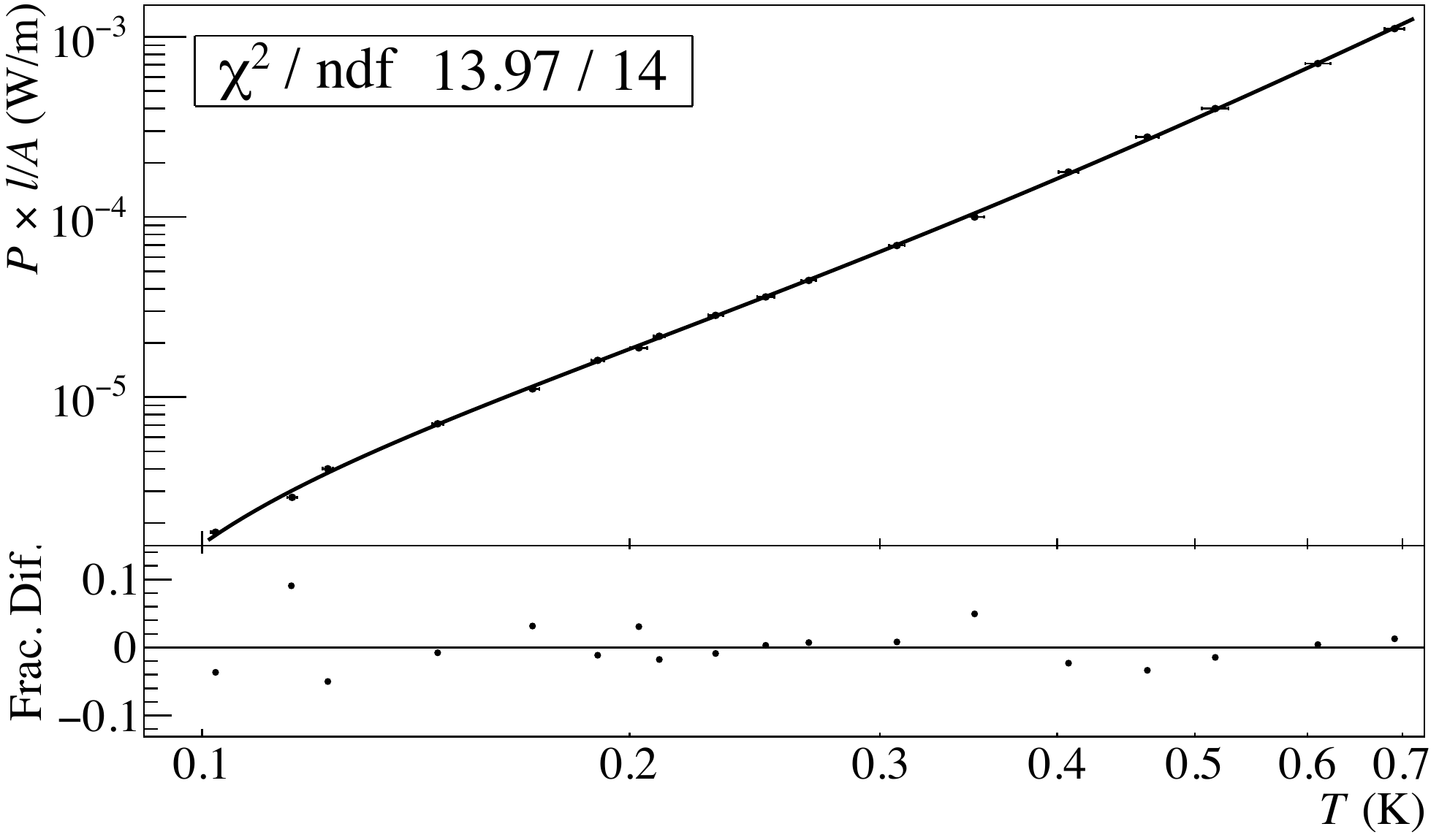}\\
\captionof{figure}{\footnotesize{Alumina sample D data with a Structure Scattering Model double power law fit (Equation 9). Fractional differences between fit and measured values are plotted below the main graph. This fit demonstrates superior agreement compared to the Equation 5 fit (Figure 6).}}\label{pinki}
}
\vspace{2mm}

\noindent dimensions of the solid in question [19]. Under the Debye approximation, so too is the phonon frequency $\omega$, hence the phonon dispersion relation gives $l(\omega) = \xi \omega ^{-n}$, with $\xi$ and $n$ denoting constants. The form of $l(\omega)$ makes our expression for thermal conductivity at low temperatures

\begin{equation}
K(T) = \frac{3\xi Nk_{\mathrm{B}}T^{3-n}}{2\pi \theta _{\mathrm{D}}^3} \int_0^{\infty}\frac{x^{4-n}e^x}{(e^x-1)^2}dx,
\end{equation}

\noindent where $x = \hbar \omega / k_{\mathrm{B}}T$, $N$ is the atomic number density, and $\theta _{\mathrm{D}}$ the Debye temperature [20].  Thus,
\begin{equation}
K(T) \propto T^{3-n}.
\end{equation}
Touloukian explains that imperfections can lead to a wide array values for $n$ that are case-specific [20]. Thus, considering the variability between alumina samples that results from the manufacturing process, we are not surprised to observe the aforementioned range of temperature dependencies above 0.3 K.

{\centering
\includegraphics[width=\columnwidth]{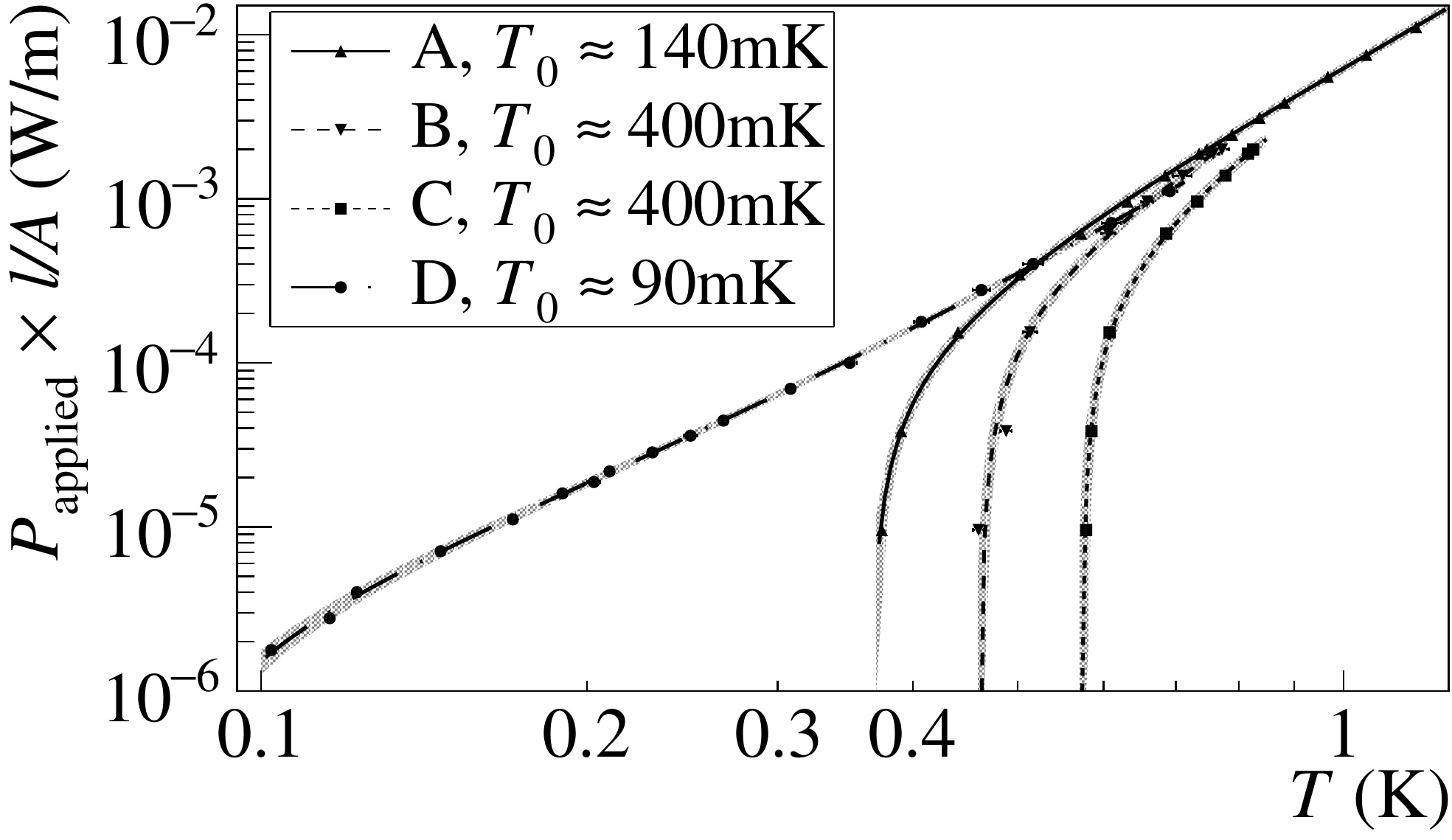}\\
\captionof{figure}{\footnotesize{Alumina measurements: applied heater power scaled by sample length $l$ and cross sectional area $A$ as a function of heated end temperature. Error bars indicate statistical uncertainty. Bands represent systematics. Letters in legend refer to samples.}}\label{pinki}
}
\vspace{2mm}

The behavior seen in Sample D is consistent with the Structure Scattering Model [21]. This involves the process of elastic scattering of phonons in semicrystalline solids due to imperfections and defects that exist on the microscopic level [22]. Scatterings can be both long and short range. The presence of long range correlations alone yields a standard ``crystalline'' $T^3$ dependence of thermal conductivity on temperature, while short range correlations yield an approximately linear term.  These correlation ranges may coexist within semicrystalline solids, resulting in an additive effect [21][22]:
\begin{equation}
K(T) = c_1 T + c_3 T^3,
\end{equation}

\noindent for some constants $c_1$ and $c_3$. In our fit, we fix the higher power value to four after noting that higher temperature measurements follow the Debye model well. Doing so reduces the number of degrees of freedom by one, which is advantageous given that the data set contains only eighteen points. We find that allowing the higher power value to float results in far larger errors on the fit parameters.

The literature exhibits variability in measured thermal conductivities of alumina. Berman reports the thermal conductivity of sintered Al$_2$O$_3$ to be proportional to $T^{2.8}$ at 2 K, attributing such behavior to the scattering of phonons off crystallite boundaries [7].

\vfill
\columnbreak

{\centering
\includegraphics[width=\columnwidth]{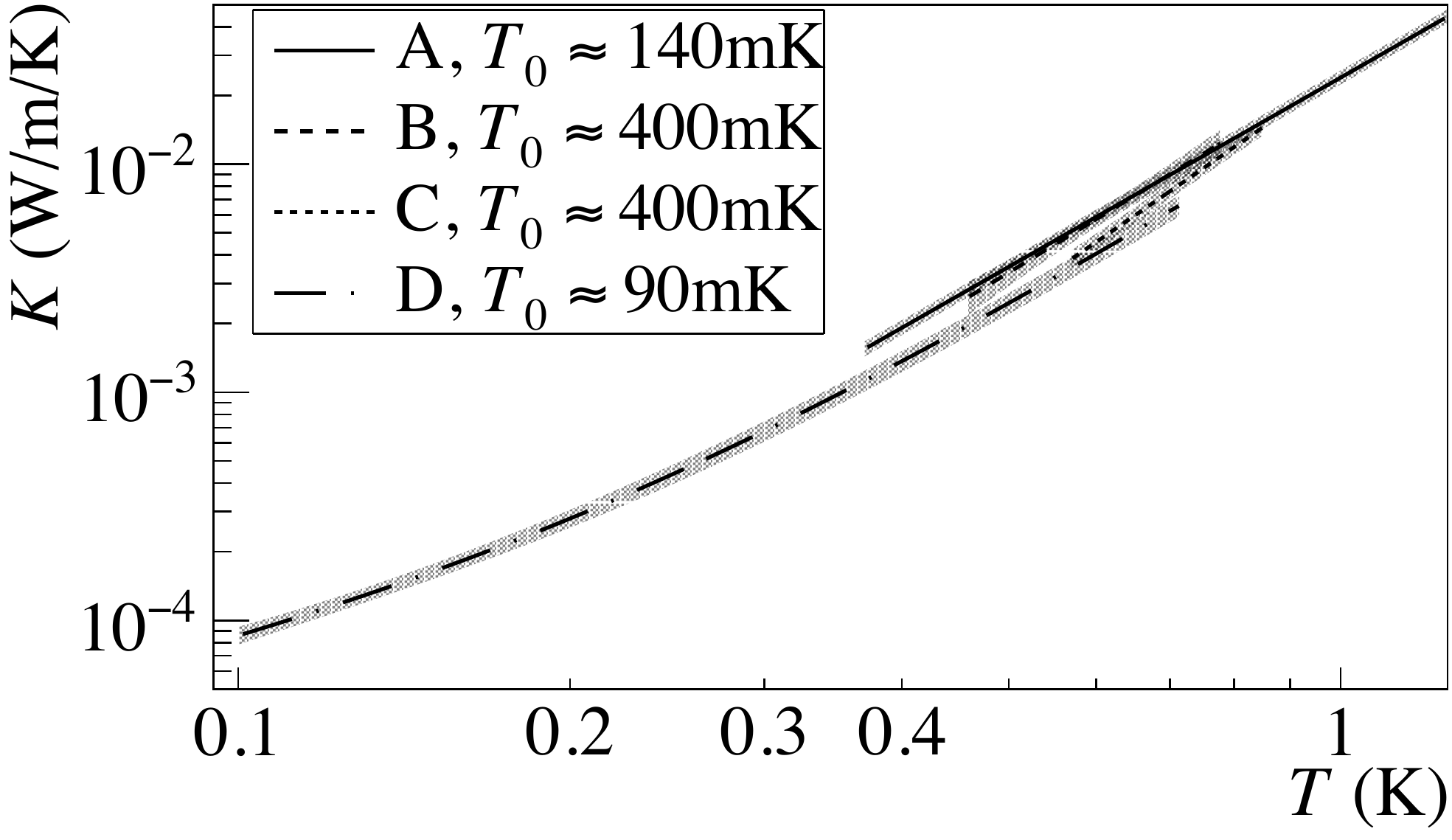}\\
\captionof{figure}{\footnotesize{Alumina thermal conductivity determined by differentiation of scaled $P_{\mathrm{applied}}$ fits in Figure 8. Shaded bands represent combined propagated statistical and systematic uncertainty. Letters in legend refer to samples.}}\label{pinki}
}
\vspace{2mm}

Alterovitz et al. measure two sintered Al$_2$O$_3$ samples yielding $T^{1.25}$ and $T^{1.4}$ dependencies between 10 and 40 K. They suspect that their samples were partially amorphous in light of the $T^{\leqslant 2}$ dependence of other amorphous materials measured by R.B. Stephens [8][23].

Comparing our data to these results, we note that power dependencies are not inconsistent with the linear and cubic phases that we see, though varied and not in precisely the same temperature regimes. In the overlapping region near 1 K, the magnitude of our measured thermal conductivity of alumina ($\sim 1 \times 10^{-2}$ W/m/K) is about an order of magnitude lower than [8] ($\sim 1 \times 10^{-1}$ W/m/K). However, at higher temperatures, where studies have been more plentiful, we do observe a fairly significant spread between measurements. For both power dependence and absolute value, fabrication differences can result in great variations from sample to sample.

It is worth mentioning that we have also verified the efficacy of the assumptions on the effective area of the heat flow using a 2-D thermal model for the samples. A 2-D thermal model suffices, considering the cylindrical symmetry for most of our samples. We used the measured thermal conductivity in our 2-D thermal model to estimate the temperature gradient in the sample, for a given applied power. As can be seen from Figure A.2, the modeled data matches

\end{multicols}
\begin{table}[h]
\centering
{\def\arraystretch{1.5}\tabcolsep=10pt
\scalebox{0.74}
{

\begin{tabular}{c c c c c c c}
\hline
Sample & Temperature Range (K) & $\mu$ (W/m/K) & $\nu$ & $\lambda$ (W/m/K) & $\gamma$ & corr[$\mu$,$\nu$] \\
\hline
A & 0.37--1.30 & (2.40 $\pm$ 0.16)$\times10^{-2}$ & 2.76 $\pm$ 0.03 & - & - & 0.014 \\
B & 0.46--0.78 & (2.59 $\pm$ 0.67)$\times10^{-2}$ & 2.95 $\pm$ 0.51 & - & - & 0.954 \\
C & 0.57--0.85 & (2.49 $\pm$ 0.24)$\times10^{-2}$ & 3.33 $\pm$ 0.18 & - & - & 0.633 \\
D & 0.10--0.72 & (1.65 $\pm$ 0.18)$\times10^{-2}$ & 3 (fixed) & (8.59 $\pm$ 3.81)$\times10^{-4}$ & 1.09 $\pm$ 0.22 & \textit{see matrix A.3}\\ \hline
\end{tabular}
}
}
\captionsetup{justification=centering}
\caption{\footnotesize{Results for alumina thermal conductivity: $K(T) = \mu (T/1 \; \mathrm{K})^{\nu}$ (A, B, C) or $K(T) = \mu (T/1 \; \mathrm{K})^{3} + \lambda (T/1 \; \mathrm{K})^{\gamma}$ (D).}}
\end{table}

\begin{multicols}{2}

\noindent extremely well with the experimental data for alumina sample D, which had the largest cross-sectional area among all the samples. The error bars on the 2-D thermal model comes from the uncertainty in the measured thermal conductivity. A short description about the 2-D thermal model can be found in Appendix 7.3.

\section{Conclusion}

We measure the thermal conductivity of PTFE in the 0.17--0.43 K range. The average result is $K_{\mathrm{w}}(T) = (5.0 \pm 0.8)\times 10^{-3} (T/1 \; \mathrm{K})^{1.83\pm 0.11}$ W/m/K. This measurement is consistent with existing results and validates our experimental techniques. We measure the thermal conductivity of alumina in the $\sim$ 0.1--1.3 K range, over which the value varies from $\sim9\times  10^{-5}$ to $\sim4 \times 10^{-2}$ W/m/K. The temperature dependence of thermal conductivity is cubic, with a smaller linear contribution coming out below $\sim$ 0.3 K, consistent with microcrystalline structure. In the vicinity of 0.1 K, we find our Alumina to be slightly less conductive than PTFE. This ceramic is widely available, and, compared to PTFE, is very rigid and solid, allowing any structural parts to be made smaller to significantly reduce heat capacity. In the process of our measurements, we find it to be much more compatible than PTFE with copper in terms of thermal contraction/expansion---we are able to hold alumina in copper clamps, which fail for PTFE due to differential contraction. As such, it appears to be a good candidate for use as a weak link in future cryogenic bolometers. We plan to test this with prototype detectors in our cryostat.

\section{Acknowledgments}

The authors thank our laboratory colleagues G. Benato, T. O'Donnell, J. L. Ouellet, B. Schmidt  and R. Hennings-Yeomans; and J. Kant and J. Wallig for their mechanical and engineering support. Funding: This work was supported by the National Science Foundation [grant numbers NSF-PHY-0902171, NSF-PHY-1314881]; and the United States Department of Energy [grant number DE-FG02-00ER41138].

\section{References}

\begin{itemize}

\item[1] Enss C (Ed). Cryogenic particle detection. Top Appl Phys. 99. 2005.

\item[2] CUORE Collaboration, Alduino C, Alfonso K, Artusa DR, Avignone FT, Azzolini O. et al. CUORE-0 detector: design, construction and operation. JINST 11. 2016. P07009.

\item[3] Cuoricino Collaboration, Arnaboldi C, Artusa DR, Avignone FT, Bandac I, Barucci M, Beeman JW. et al. Results from a search for the 0 neutrnio beta beta-decay of Te-130. Phys Rev C. 2008. 035503.

\item[4] Scott T, Giles M. Dislocation scattering in Teflon at low temperatures. Phys Rev Lett 29. 1972. 642--3. 

\item[5] Marquardt ED, Le JP. Radebaugh R, Cryogenic material properties database. NIST. Boulder, CO 80303.

\item[6] Vignati M. Model of the response function of CUORE bolometers. PhD diss. Sapienza Universit\`{a} di Roma. 2009.

\item[7] Berman R. The thermal conductivity of some polycrystalline solids at low temperatures. Proc Phys Soc A. 65. 1952. 1029--1040.

\item[8] Alterovitz S, Deutscher G, Gershenson M. Heat capacity and thermal conductivity of sintered Al$_2$O$_3$ at low temperatures by the heat pulse technique. J Appl Phys. 46. 1975. 3637--3643.

\item[9] Ventura G, Risegari L. The art of cryogenics: low-temperature experimental techniques. Elsevier, Oxford. 2008.

\item[10] Swartz ET, Pohl RO. Thermal boundary resistance. Rev Mod Phys. 61. 1989. 605-668. 

\item[11] \sloppy McMaster-Carr$^{\text{\tiny{\textregistered}}}$. Raw materials. 
\burl{https://www.mcmaster.com/\# raw-materials/=14xkupj}. Accessed 7 November 2016.

\item[12] Peroni I, Gottardi E, Peruzzi A, Ponti G, Ventura G. Nuclear Physics B (Proc Suppl). 78. 1999. 573--5. 

\item[13] Olson JR. Cryogenics. 33. 1993. 729--731. 

\item[14] Cheeke J. The kapitza resistance and heat transfer at low temperatures. J de Physique Colleques. 31 (C3). 1970. C3-129--C3-136. 

\item[15] Gubrud MA. Scanning tunneling microscopy at millikelvin temperatures: design and construction. PhD diss. University of Maryland, College Park. 2010.

\item[16] Anderson AC, Reese W, Wheatley JC. Thermal conductivity of some amorphous dielectric solids below 1 $^{\circ}$K. Rev Sci Instr. 34. 1386.

\item[17] Anderson PW, Halperin BI, Varma CM. Anomalous low-temperature thermal properties of glasses and spin glasses. Philosophical Magazine. 25. 1972. 1--9.

\item[18] Phillips WA. Tunneling states in amorphous solids. J Low Temperature Phys. 7. 1972.

\item[19] Kittel C. Introduction to solid state physics. John Wiley \& Sons, Inc. 5. 2005. 105--129.

\item[20] Touloukian YS, Powell RW, Ho CY, Klemens PG. Thermophysical properties of matter vol. 2: thermal conductivity, nonmetallic solids. Plenum Press, New York. 1970.

\item[21] Finlayson DM, Mason PJ. Structure scattering and the thermal conductivity of a semicrystalline polymer. J Phys C: Solid State Phys. 18. 1984. 1791--1802.  

\item[22] Morgan GJ, Smith D. Thermal conduction in glasses and polymers at low temperatures. J Phys C: Solid State Phys. 7. 1973. 649--664. 

\item [23] Stephens RB. Low-temperature specific heat and thermal conductivity of noncrystalline dielectric solids. Phys Rev B. 8. 1973. 2896.

\item [24] \sloppy MathWorks$^{\text{\tiny{\textregistered}}}$. Documentation. \burl{https://www.mathworks.com/help/pde/heat-transfer-and-diffusion-equations.html}. Accessed 16 May 2017.

\end{itemize}

\end{multicols}

\section{Appendix}

\subsection{Applied heater power fit results}
\setcounter{table}{0}
\renewcommand{\thetable}{A.\arabic{table}}

\begin{table}[h]
\centering
{\def\arraystretch{1.5}\tabcolsep=10pt
\scalebox{0.65}
{
\begin{tabular}{c c c c c c}
\hline
Sample & Temperature Range (K) & $a$ (W/m) & $b$ & $c$ (W/m) & corr[$a$,$b$] \\ 
\hline
1 & 0.18--0.26 & (1.52 $\pm$ 0.29)$\times 10^{-3}$ & 2.70 $\pm$ 0.17 & (-1.57 $\pm$ 0.15)$\times 10^{-5}$ & 0.983 \\ 
1 & 0.27--0.36 & (2.07 $\pm$ 0.35)$\times 10^{-3}$ & 2.99 $\pm$ 0.20 & (-4.15 $\pm$ 0.42)$\times 10^{-5}$ & 0.977 \\
1 & 0.37--0.43 & (2.42 $\pm$ 0.60)$\times 10^{-3}$ & 3.05 $\pm$ 0.41 & (-1.12 $\pm$ 0.19)$\times 10^{-4}$ & 0.989 \\ 
2 & 0.17--0.21 & (1.77 $\pm$ 0.50)$\times 10^{-3}$ & 2.82 $\pm$ 0.22 & (-0.84 $\pm$ 0.12)$\times 10^{-5}$ & 0.990 \\ \hline

\end{tabular}
}
}
\captionsetup{justification=centering}
\caption{\footnotesize{Results for the PTFE scaled power fits: $P_{\mathrm{applied}}(T)\times l/A = a(T/1 \; \mathrm{K})^b + c$.}}
\end{table}

\begin{table}[h]
\centering
{\def\arraystretch{1.5}\tabcolsep=10pt
\scalebox{0.65}
{
\begin{tabular}{c c c c c c c c}
\hline
Sample & Temperature Range (K) & $a$ (W/m) & $b$ & $c$ (W/m) & $r$ (W/m) & $s$ & corr[$a$,$b$]    \\ 
\hline
A & 0.37--1.30 & (6.38 $\pm$ 0.43)$\times10^{-3}$ & 3.76 $\pm$ 0.03 & (-1.5 $\pm$  0.1)$\times10^{-4}$ & - & - & -0.105 \\ 

B & 0.46--0.78 & (6.55 $\pm$ 0.93)$\times10^{-3}$ & 3.95 $\pm$ 0.51 & (-3.1 $\pm$ 0.9)$\times10^{-4}$ & - & - & 0.838 \\ 

C & 0.57--0.85 & (5.76 $\pm$ 0.44)$\times10^{-3}$ & 4.33 $\pm$ 0.18 & (-5.2 $\pm$ 0.5)$\times10^{-4}$ & - & - & 0.240 \\  

D & 0.10--0.72 & (4.12 $\pm$ 0.45)$\times10^{-3}$ & 4 (fixed) & (-2.2 $\pm$ 0.6)$\times10^{-6}$ & (4.1 $\pm$ 1.4)$\times10^{-4}$ & 2.09 $\pm$ 0.22 & \textit{see matrix A.4}  \\ \hline

\end{tabular}
}
}
\captionsetup{justification=centering}
\caption{\footnotesize{Results for the alumina scaled power fits: $P_{\mathrm{applied}}(T)\times l/A = a(T/1 \; \mathrm{K})^b + c$ (A, B, C) or $P_{\mathrm{applied}}(T)\times l/A = a(T/1 \; \mathrm{K})^4 + r(T/1 \; \mathrm{K})^s + c$ (D).}}
\end{table}

\subsection{Alumina sample D correlation matrices}

\begin{center}

$\begin{pmatrix}
    \mathrm{corr}[\mu,\mu] & \mathrm{corr}[\mu,\lambda] & \mathrm{corr}[\mu,\gamma] \\
    \mathrm{corr}[\lambda,\mu] & \mathrm{corr}[\lambda,\lambda] & \mathrm{corr}[\lambda,\gamma] \\
    \mathrm{corr}[\gamma,\mu] & \mathrm{corr}[\gamma,\lambda] & \mathrm{corr}[\gamma,\gamma] \\
  \end{pmatrix}
  =
  \begin{pmatrix}
    \phantom{-}1.000 & -0.003 & -0.004\\
    -0.003 & \phantom{-}1.000 & \phantom{-}0.979 \\
    -0.004 & \phantom{-}0.979 & \phantom{-}1.000 \\
  \end{pmatrix}$\\

\vspace{2mm}
Matrix A.3: {\footnotesize{Alumina sample D thermal conductivity  $K(T) = \mu(T/1 \; \mathrm{K})^3 + \lambda(T/1 \; \mathrm{K})^{\gamma}$) parameter correlations. Parameter values and uncertainties are given in Table A.2. $\nu$ is fixed, and correlations with it are trivially zero.}}

\vspace{4mm}

$\begin{pmatrix}
    \mathrm{corr}[a,a] & \mathrm{corr}[a,r] & \mathrm{corr}[a,s] \\
    \mathrm{corr}[r,a] & \mathrm{corr}[r,r] & \mathrm{corr}[r,s] \\
    \mathrm{corr}[s,a] & \mathrm{corr}[s,r] & \mathrm{corr}[s,s] \\
  \end{pmatrix}
  =
  \begin{pmatrix}
    \phantom{-}1.000 & -0.812 & -0.861 \\
    -0.812 & \phantom{-}1.000 & \phantom{-}0.965 \\
    -0.861 & \phantom{-}0.965 & \phantom{-}1.000 \\
  \end{pmatrix}$\\

\vspace{2mm}
Matrix A.4: {\footnotesize{Alumina sample D scaled power fit $P_{\mathrm{applied}}(T)\times l/A = a(T/1 \; \mathrm{K})^4 + r(T/1 \; \mathrm{K})^s + c$ parameter correlations. Parameter values and uncertainties are given in Table A.2. $b$ is fixed, and correlations with it are trivially zero. Correlations with $c$ do not affect thermal conductivity calculation.}}

\end{center}

\subsection{2-D thermal model}

The steady state heat transfer in a circular cylindrical rod, in the absence of internal heat source, can be described by the following equation 
\begin{equation}
\nabla \cdot (K \nabla T) = 0,
\end{equation}
where $K$ is the thermal conductivity of the rod. Since the sample is axisymmetric, it is convenient to write the above expression in a cylindrical form
\begin{equation}
\frac{1}{r}\frac{\partial}{\partial r}\left(Kr\frac{\partial T}{\partial r}\right)- \frac{\partial}{\partial z}\left(K\frac{\partial T}{\partial z}\right)=0.
\end{equation}
The above equation allows us to analyze the 3-D axisymmetric problem using a 2-D model. A MATLAB routine, the Partial Differential Equation Toolbox$^{\text{\tiny{TM}}}$ [24], is used to analyze the model with appropriate boundary conditions. These include keeping one side of the rod at a constant temperature, while continuously adding heat from the other side. We assume that the heat transfer at the outer boundary takes place solely due to radiation exchange. The ambient temperature is fixed at 0.1 K for the same. The output of this routine for alumina sample D is displayed below in Figure A.1. Figure A.2 shows the applied power against temperature, as obtained by the thermal model for alumina sample D, overlayed with the data and fit from Figure 8, where we assume 1-dimensional  heat flow. The 2-D model agrees with measurements within $5\%$.

\begin{center}
\includegraphics[scale=0.4]{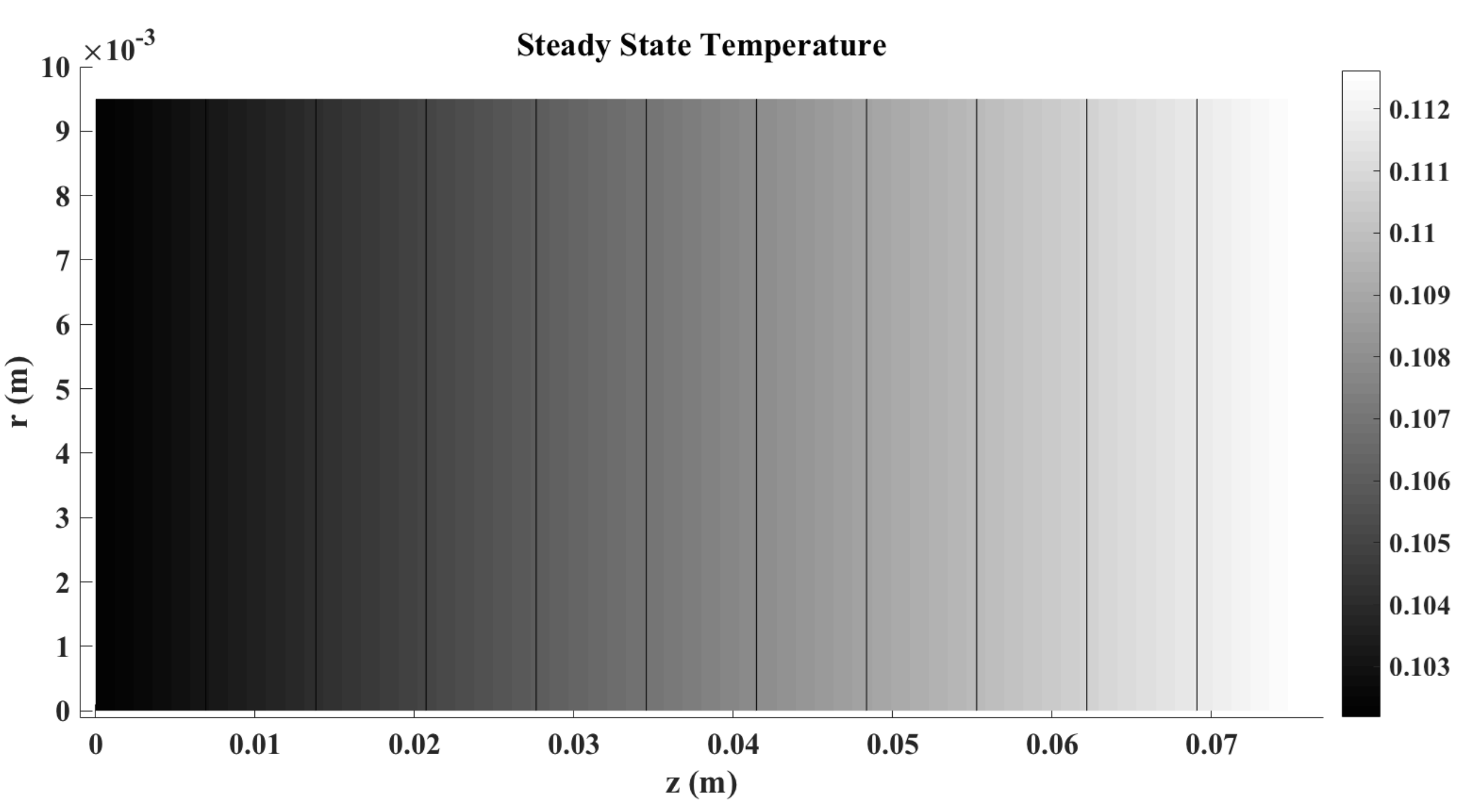}\\
Figure A.1: {\footnotesize{Temperature gradient along the alumina sample D at 100 mK. The edge at $z$ = 0 m was kept at a constant temperature while a constant heat power of $P_{\mathrm{applied}}$ = 4.5 nW was used as a boundary condition on the other end ($z$ = 0.076 m). Numerical solution to Equation 15 was obtained through use of the Partial Differential Equation Toolbox$^{\text{\tiny{TM}}}$ and using the measured $K$ for the sample. The vertical lines denote isotherms and clearly shows an uniform heat flow along the sample.}}
\end{center}
\vspace{2mm}
\begin{center}
\includegraphics[scale=0.4]{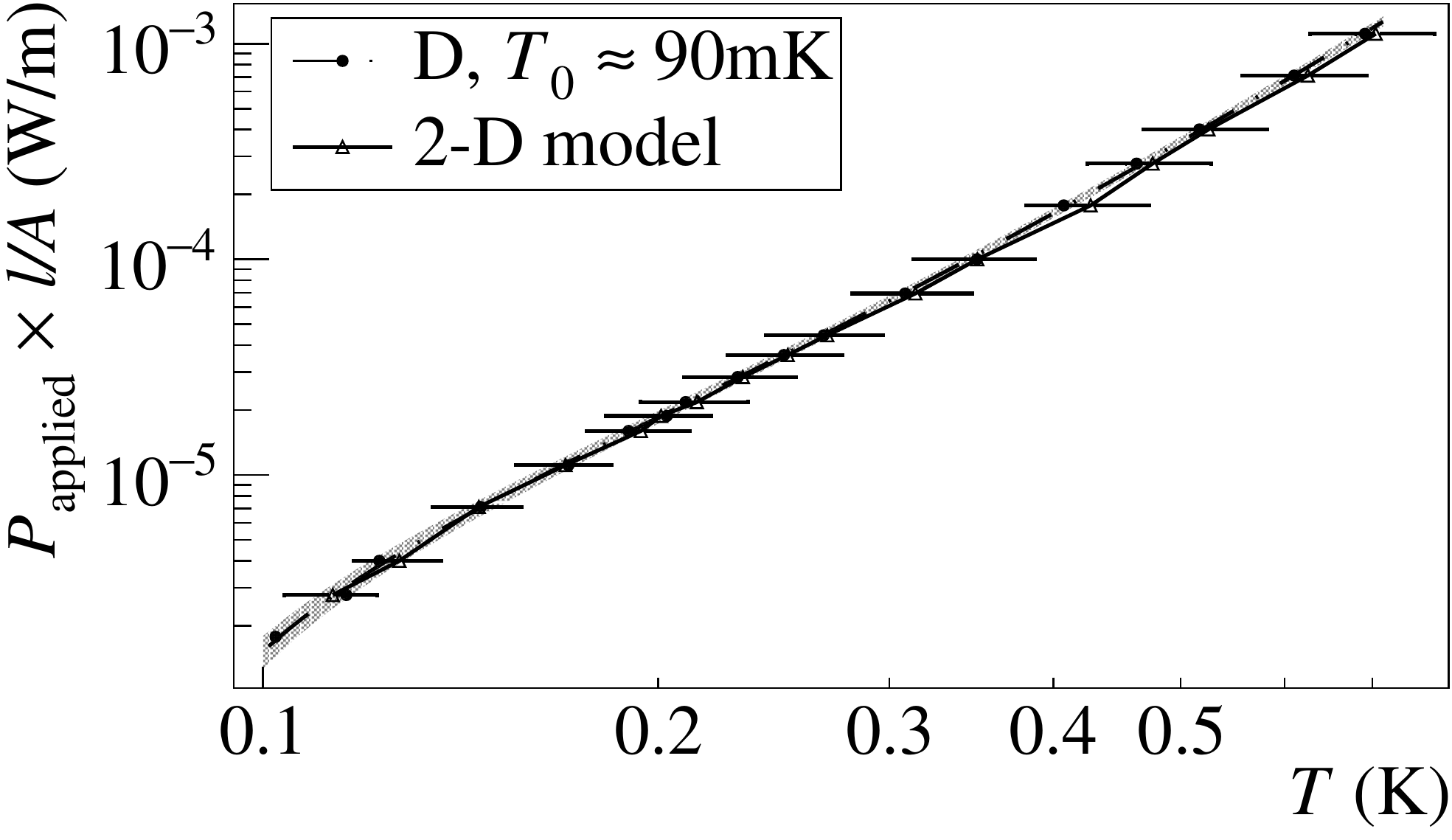}\\
Figure A.2: {\footnotesize{Plot of applied power versus temperature as obtained from the 2-D thermal model and overlaid on the experimental data points for alumina sample D. The horizontal error bars on the 2-D model originates from propagation of the uncertainty in the measured thermal conductivity that was used in the thermal model.}}
\end{center}

\end{document}